\documentclass{article}
\usepackage{amsmath,graphicx}
\usepackage[preprint]{spconf}
\usepackage{csquotes}

\copyrightnotice{\copyright 2018 IEEE. Personal use of this material is permitted. Permission from IEEE must be obtained for all other uses. }

\title{ Speech Prediction using an Adaptive Recurrent Neural Network \\ with Application to Packet Loss Concealment }
\name{Reza Lotfidereshgi, Philippe Gournay}
\address{Speech and Audio Research Group\\
Universit\'e de Sherbrooke\\
Sherbrooke (Qu\'ebec) J1K 2R1 Canada\\\\}
\begin{document}
%
\maketitle
\begin{abstract}

This paper proposes a novel approach for speech signal prediction based on a recurrent neural network (RNN). Unlike existing RNN-based predictors, which operate on parametric features and are trained offline on a large collection of such features, the proposed predictor operates directly on speech samples and is trained online on the recent past of the speech signal. Optionally, the network can be pre-trained offline to speed-up convergence at start-up. The proposed predictor is a single end-to-end network that captures all sorts of dependencies between samples, and therefore has the potential to outperform classical linear/non-linear and short-term/long-term speech predictor structures. We apply it to the packet loss concealment (PLC) problem and show that it outperforms the standard ITU G.711 Appendix I PLC technique.


%
%

\end{abstract}

\section{Introduction}
\label{sec:intro}

Consecutive samples or blocks of natural signals are usually correlated with one another. In the case of speech signals \cite{clark2008introduction}, most of this correlation comes from the human speech production mechanism, which is by no means memoryless (due to the inertia of the vocal cords and articulators, and to resonances in the vocal tract). In addition to the particular anatomy of the speaker, the limited set of phonemes and words that compose the language he or she speaks (linguistics) and the specific message he or she wants to share (semantics) both introduce further correlation. Successful speech processing applications, including speech compression, recognition and synthesis, make extensive use of these correlations.

In classical speech signal processing, specific structures called predictors are normally used to capture and make use of these correlations. Linear Predictive Coding (LPC) for example relies on various types of linear predictors \cite{makhoul1975linear}. Short-term predictors, which operate at the sample scale (one millisecond or less), handle correlations between nearby samples. Pitch predictors, which operate at a larger time scale (one pitch period, typically 2.5 ms to 20 ms), deal with longer-term correlations. To increase performance, nonlinear predictors (for example based on Volterra or Wiener series) are sometimes used to capture more subtle correlations \cite{thyssen1994non}. These different predictors are generally combined in a cascade, each one taking care of its own type of correlation.

Recently, new machine learning and artificial intelligence tools have been developed to capture correlations in sequential data such as text and speech signals \cite {lecun2015deep}. Hidden Markov Models (HMMs) and Deep Neural Networks (DNNs) \cite{hinton2012deep}, which are very efficient at unveiling statistical dependencies for the former and nonlinear dependencies for the latter, are two examples of such tools. Several speech processing applications, in particular speech and speaker recognition and speech synthesis, have made sudden and considerable progress since they were introduced \cite{lecun2015deep}.

These modern tools are often treated as black boxes in the sense that the painstaking manual tuning that characterised most classical tools has been replaced by an automatic, thus effortless (yet computationally intensive), training on the largest possible dataset. This approach makes it possible to design very complex systems that provide greater performance. Most of them, however, still rely on a combination of manually and cleverly designed features.

In this paper, we propose a new approach for speech signal prediction based on an adaptive recurrent neural network. Compared to existing approaches, the network operates directly on speech signal samples and is actively trained on the recent past of the speech signal. To demonstrate the merits of this predictor, we apply it to the Packet Loss Concealment (PLC) problem and compare it to the ITU G.711 Appendix I standard \cite{rec1999711}.

The outline of the paper is as follows. First, the Long Short-Term Memory (LSTM) architecture of Recurrent Neural Networks (RNN), along with classical and modern techniques for Packet Loss Concealment (PLC), are briefly reviewed in section \ref{sec:prio}. The proposed speech predictor structure and its application to PLC are then described in detail in section \ref{sec:meth}. The experimental setup and  results obtained are presented in section \ref{sec:exp}. Finally, some conclusion are drawn and perspectives for future research are discussed in section \ref{sec:conc}.

\section{Relation to prior work}
\label{sec:prio}

This section presents a brief overview of the long short-term memory architecture used in the proposed approach for speech prediction, and some background information about packet loss concealment.

\subsection{Long Short-Term Memory}
\label{ssec:LSTM}

Long Short-Term Memory (LSTM) is a variation of Recurrent Neural Network (RNN) that is very efficient at solving problems related to sequential data. The concept was first proposed in 1997 \cite{hochreiter1997long} then refined over the years \cite{gers1999learning,gers2000recurrent,graves2005framewise}. An LSTM network is composed of blocks, each block containing different gates that control the flow of information. The \enquote{input} gate controls the flow of information from the input of the block to its memory. The   \enquote{forget}  gate controls the duration for which the information is kept in memory. Finally, the \enquote{output} gate controls the contribution of the memorized information to the output activation of the block. Training of LSTM networks is normally done using backpropagation through time \cite{graves2005framewise}.

Over the past years, LSTM and other structures with gated units have proven to be very successful in solving various speech-related problems. This goes from processing of text information including automatic translation \cite{cho2014learning}, to processing of acoustic signals such as speech recognition \cite{graves2013speech,graves2014towards}. Most applications to acoustic speech signals consist in classification or recognition tasks. For these applications, training the network first then operating (or testing) it makes much sense. Also, for these applications, the network usually operates on parametric features such as MFCCs or spectrograms rather than directly on input samples, because these features provides some degree of abstraction and the information lost does not really matter.

\subsection{Packet Loss Concealment}
\label{ssec:PLC}


The purpose of Packet Loss Concealment (PLC) in a Voice over Packet Network (VoPN) speech communication system is to provide a replacement for unavailable (either lost or overly delayed) speech packets \cite{perkins1998survey}. Most, and possibly all, conventional (i.e. signal processing-based) PLC techniques rely exclusively on the most recent past and sometimes on the most immediate future of the speech signal. Conventional PLC techniques that operate directly in the signal domain simply extrapolate or interpolate from one or two pitch periods before and after the packet loss. Model-based or decoder-based conventional PLC techniques rely on parametric features or speech coding parameters that represent one or two frames of speech signal (one frame being typically 10 to 30 ms long). All these techniques have been engineered, or manually designed, based on the general properties of the speech signal. In particular, the short time-horizon that is used stems from the short decorrelation time that characterizes speech signals.

In the last decade, some more modern (i.e. machine learning or artificial intelligence-based) PLC techniques have been proposed. In reference \cite{rodbro2006hidden}, a statistical Hidden Markov Model (HMM) is used to drive a sinusoidal analysis/synthesis model of the speech signal. In reference \cite{lee2016packet}, a Deep Neural Network (DNN) is used to regenerate the log-power spectrum and phases of the missing frame. The DNN is first trained on a large set of spectral features. Then, in the reconstruction stage, it is fed with the spectral features of the previous frames. To our best knowledge, a modern PLC technique that that is not pre-trained and that operates directly in the signal domain has  yet to be proposed.

%



\begin{figure}[t]

  \centerline{\includegraphics[width=8.5cm]{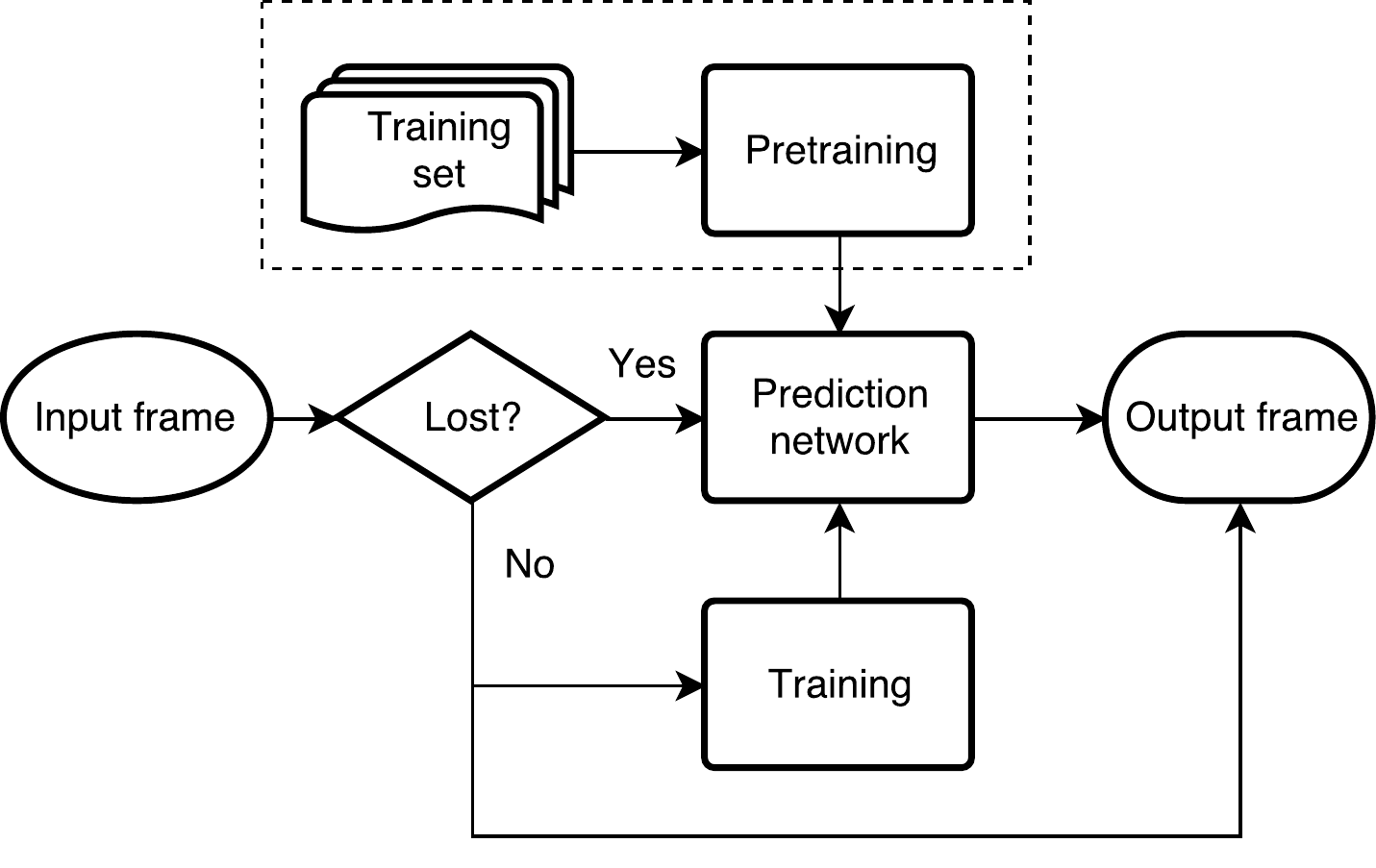}}
\caption{Flowchart of the proposed PLC algorithm.}
\label{fig:flow}
\end{figure}

\section{Proposed method}
\label{sec:meth}

This section presents the general structure of the proposed predictor then gives details about offline pretraining, online training and prediction.

\subsection{General structure}
\label{sec:gene}
The proposed method for speech prediction and packet loss concealment is illustrated in Fig.\ref{fig:flow}. First, as indicated in the dotted-line box, the prediction network can be either randomly initialized or pretrained offline on a set of speech signals. This pretraining is optional but has the advantage of speeding-up convergence of the network at start-up. Then, depending on whether the input speech frame is available or not, the decision is made to simply copy it to the output and train (i.e. adapt or update) the prediction network, or to generate a replacement for the lost frame using the prediction network.

The prediction network operates directly on raw speech samples, without any feature extraction. This is the current trend in speech processing using neural networks \cite{trigeorgis2016adieu}, mostly because no set of parametric feature has been found so far that is complete enough to capture all dimensions of the speech signal (including speaker identity, emotional state and acoustic background ambiance).

During our first experiments, we considered two options to feed the network: either one single sample at every time step, or with a sliding window of consecutive samples at every time step (with a window shift of one sample). We used the latter approach for two reasons. First, because it allows the network to learn both an internal representation of the speech signal and its evolution through time. Then, because the former approach has proven to produce less stable results especially when reconstructing lost frames (the error in a predicted sample seems to propagate much easier in the single input network).

The prediction network itself is composed of several layers of LSTM blocks. In the experiments described in section \ref{sec:exp}, one of the most commonly used LSTM architecture called Vanilla LTSM \cite{greff2017lstm} is used. LSTM states are reset to zero between batches (stateless mode). Peephole connections were not implemented because, according to \cite{greff2017lstm}, they do not bring much benefit for prediction. The Mean Squared Error (MSE) between input samples and predicted samples is used as objective training criterion.

\begin{figure}[t]

  \centerline{\includegraphics[width=8.5cm]{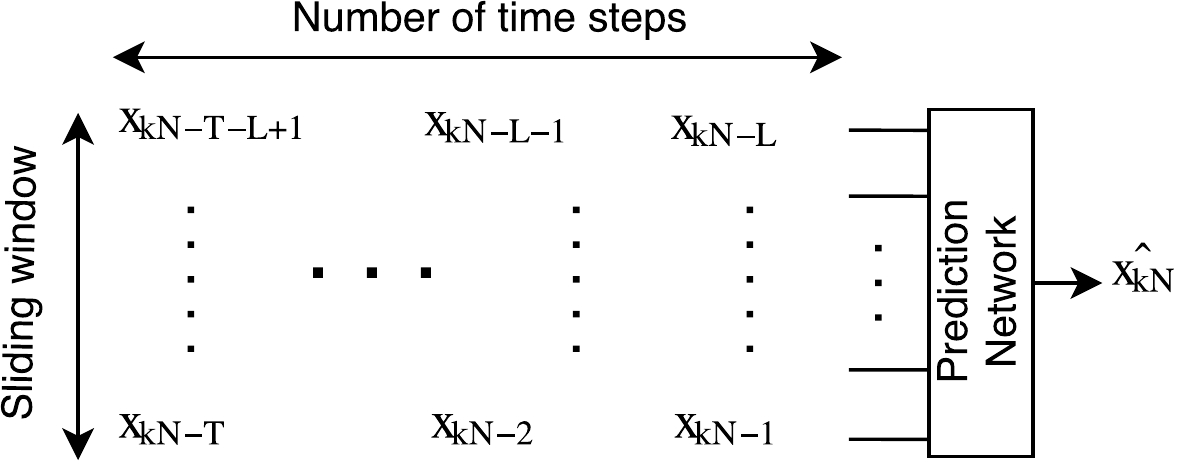}}
\caption{Signal samples used to train the prediction network for the first sample of frame number k. N is the frame duration in samples, L is the size of the sliding window, and T is the number of time steps}
\label{fig:inputs}
\end{figure}

\subsection{Offline pretraining}
\label{sec:pret}

Optional offline pretraining on a set of speech signals can be used to initialize weights and biases in the network. In the experiments described in section \ref{sec:exp}, this is done using the Adam optimizer \cite{kingma2014adam} and minibatches of 80 samples (10 ms at the 8 kHz sampling rate).

This pretraining alone is not enough to provide good prediction performance. Speech is both an exceptionally diverse and extremely dynamic signal. It is therefore difficult for a single network to follow continuously and accurately its slightest local variations. Training and executing a single giant network to do so does not seem practical either. Keeping track of the evolution of the signal while maintaining a reasonable complexity requires using a smaller network and training it on a more representative set of signals. This is why we use online training on the recent past of the speech signal.
\subsection{Online training or prediction}
\label{sec:train}

There are two possibilities during the operation phase of the network. If the input speech frame is available, it is simply copied to the output of the system as illustrated in Fig.\ref{fig:flow}. It is also considered as a small but extremely relevant training set for the network. Multiple training passes are made over that small training set in a process that is often referred to as stochastic optimization \cite{kingma2014adam}. The samples used to train the prediction network are illustrated in Fig.\ref{fig:inputs}.

If no input frame is available, then the prediction network performs regression to predict the first sample of the output frame. This sample  is then used as input to predict the next sample of the lost frame. This process goes on until the entire lost frame has been reconstructed.

\begin{figure}[t]

  \centerline{\includegraphics[width=8.5cm]{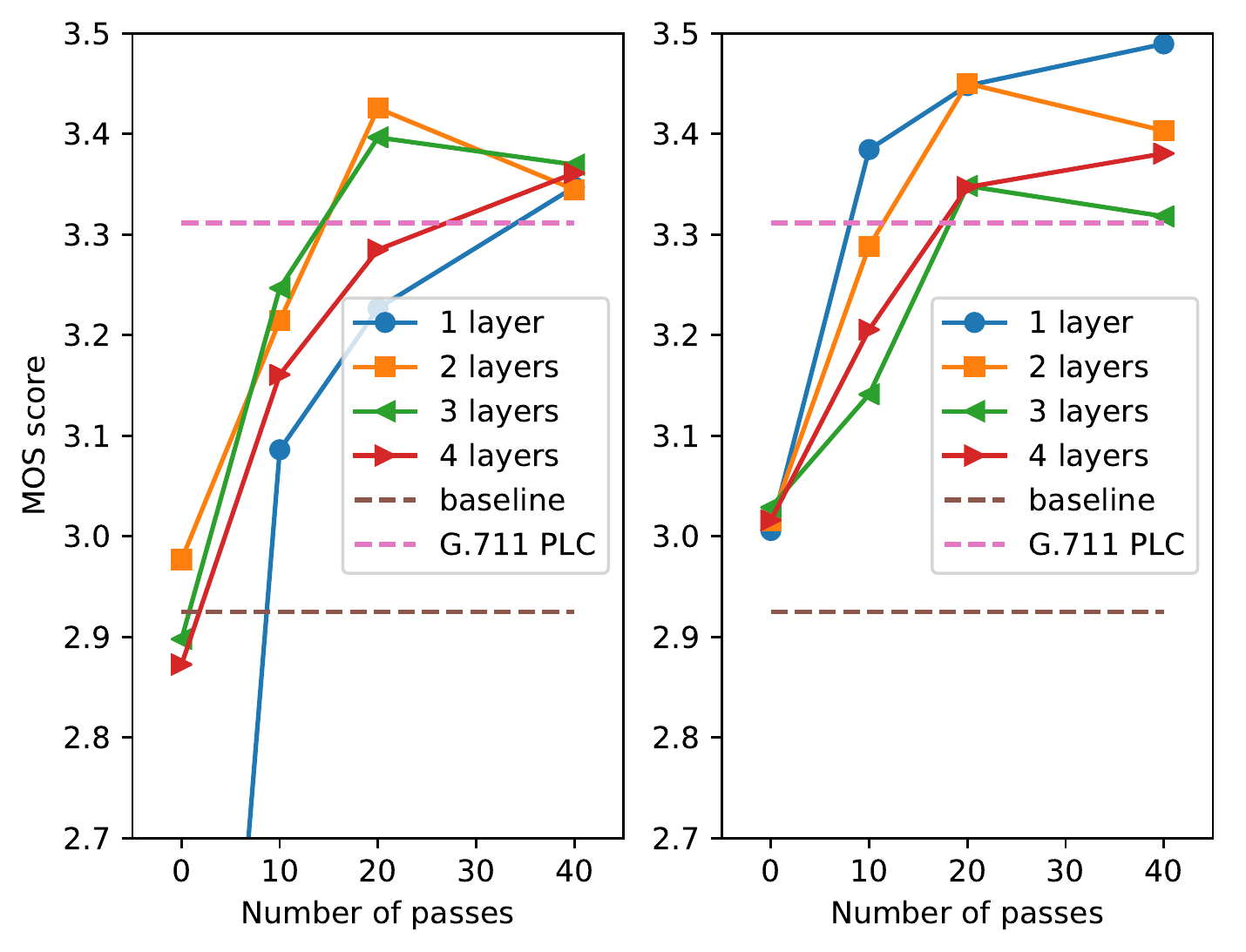}}
\caption{MOS score as a function of number of passes for different configurations of the network, for 80 time steps (left) and 160 time steps (right).}
\label{fig:tuning1}
\end{figure}

\section{EXPERIMENTS}
\label{sec:exp}

The experiments presented in this section are intended to explore a variety of configurations for the proposed speech predictor. Performance on a PLC task is studied because this is a straightforward and typical application for speech prediction. Also, the ITU-T G.711 Appendix I standard is used as a reference, not only because of its good performance but also because it is a representative example of conventional signal processing-based PLC.

\subsection{Speech material}
\label{sec:speech}
The experiments are done on a subset of the TIMIT database \cite{garofolo1993darpa}. The sampling frequency is 8000 Hz and the frame (or packet) duration is set to 10 ms. The number of neurons per layer is intentionally small to limit the complexity of online training. To reduce the extent of the experiments, the length of the sliding window is chosen to always be the same as the number of neurons per layer.

Two hundred speech files from the training subset of the TIMIT database are used for pretraining. They represent more than 50,000 frames of speech signal, for a total of 4 million of training speech samples. Since the number of neurons per layer is small, a single epoch of pretraining is performed.

Ten speech files from the testing subset of the TIMIT database are used for testing. A 10\% packet loss rate is simulated, with lost packets being evenly spaced. The test files therefore represent more than 400 lost packets, for a total of 32,000 lost, hence predicted, samples. The performance of the proposed method is evaluated in terms of Mean Opinion Score (MOS) and compared to that of the G.711 PLC algorithm. The MOS score is obtained using the Perceptual Evaluation of Speech Quality (PESQ) software tool \cite{hu2008evaluation}.

\subsection{Results}
\label{sec:results}

Fig.\ref{fig:tuning1} present the MOS as a function of the number of training passes for different numbers of LTSM layers (from 1 to 4). The left panel corresponds to 80 time steps and the right one to 160 times steps. In all cases, both the number of neurons per layer and the size of the sliding window are set to 80. The lower and upper dotted lines correspond to the MOS of a baseline condition (lost packets simply set to zero) and to the MOS of the G.711 PLC algorithm, respectively. From these results, we conclude that the network cannot predict speech efficiently after pretraining only (zero training passes). Specifically, in the case of 80 time steps, the performance of the proposed PLC system is generally below the baseline condition. Another conclusion is that better performance is obtained with 160 time steps than with 80. Also, 20 training passes generally gives good performance. Lower performance for more that 20 passes probably results from overfitting because each pass is performed on a small number of data.

\begin{figure}[t]

  \centerline{\includegraphics[width=8.5cm]{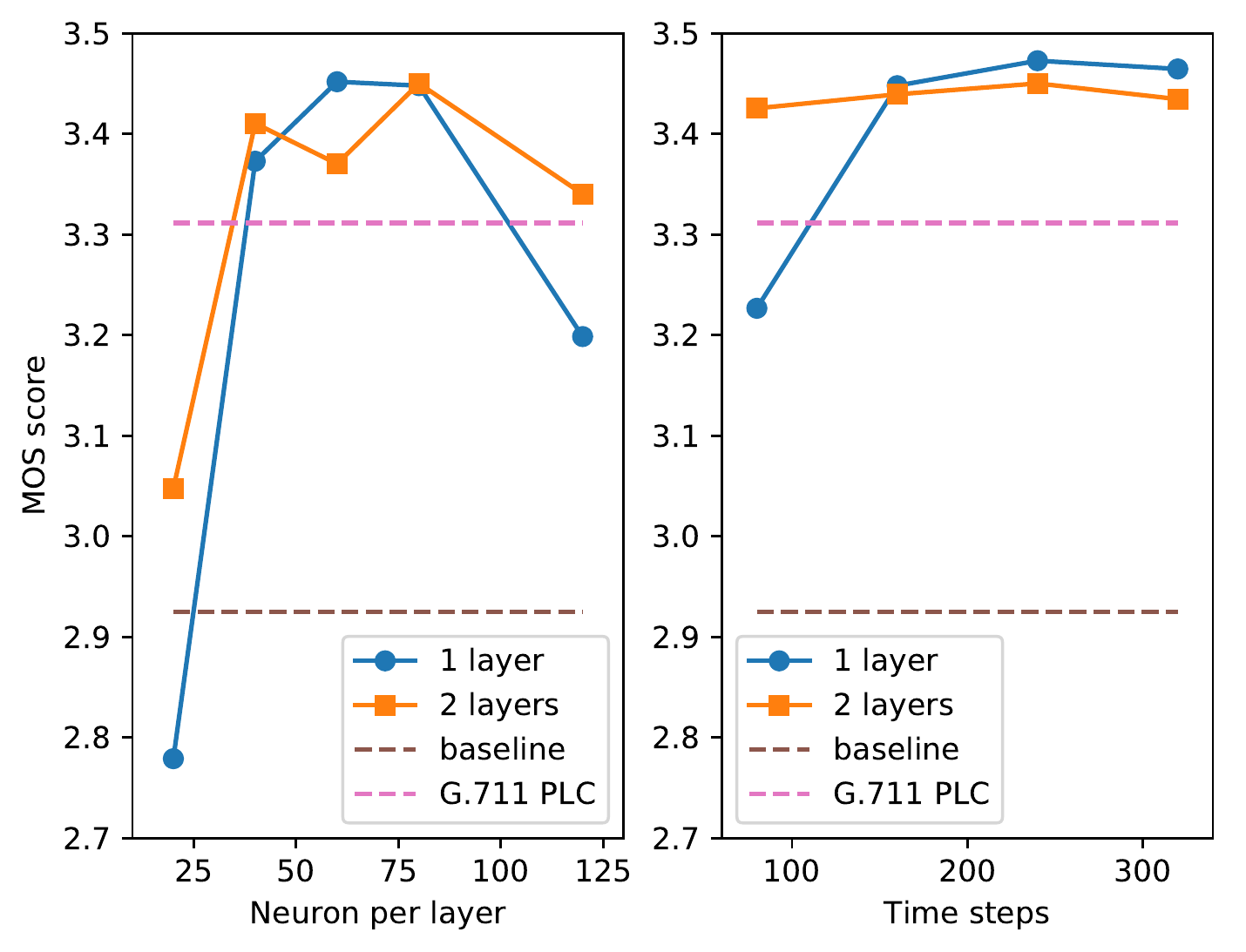}}
\caption{MOS score as a function of the number of neurons per layer (left) of the number of time steps (right).}
\label{fig:tuning2}
\end{figure}

Fig.\ref{fig:tuning2} presents the results obtained when varying the number of neurons per layers (left panel) and the number of time steps (right panel). In both cases, the number of passes is equal to 20. The first conclusion is that even a single layer of only 40 LSTM neurons performs reasonably well when online training is used. The second conclusion is that increasing the number of time steps beyond 160 does not seem to increase performance significantly.

%
%
%
%
%
%
%
%
%
%
%
%
%

\section{CONCLUSIONS}
\label{sec:conc}

In this paper, a new approach for speech signal prediction based on a recurrent neural network was proposed. The main characteristic of this approach is that the network is actively trained on the recent past of the signal. Since this recent past represents the best possible training set almost at all times, the network is able to follow closely the evolutions of the signal. Furthermore, since the local variability of the signal is limited compared to the variability of speech in general, a small network structure is effective. This, in turn, allows the network to operate directly on speech samples. And since there is no more need for a manually-design feature representation, no information is lost compared to systems that include a feature extraction step.

The proposed speech signal predictor was tested on a packet loss concealment task. With proper setting, it was shown to outperform the standard ITU-T G.711 Appendix I PLC algorithm. It is interesting to note that these good results were obtained using a completely speech-agnostic system, in the sense that no speech model nor prior information about subjective speech quality evaluation was introduced in its design. Good performance was achieved even when using a very small LSTM networks (one layer of forty neurons). This is interesting because complexity can rapidly be an issue when performing online training or output regression.

Only a limited number of LSTM configurations were tested. Different neural network configurations, different pretraining and training procedures, or even different types of neural networks may further improve performances.

Finally, the proposed predictor is a very general tool that could benefit to other applications in speech processing, and that could apply to other correlated yet highly dynamic types of data.

\newpage

\bibliographystyle{IEEEbib}
\bibliography{clark2008introduction,hu2008evaluation,rec1999711,makhoul1975linear,lecun2015deep,hinton2012deep,garofolo1993darpa,greff2017lstm,kingma2014adam,trigeorgis2016adieu,thyssen1994non,rodbro2006hidden,lee2016packet,takahashi2004perceptual,perkins1998survey,cho2014learning,recommendation1999711,graves2014towards,graves2013speech,hochreiter1997long,gers1999learning,gers2000recurrent,graves2005framewise}

\end{document}